\documentclass[aps,pra,twocolumn,groupedaddress,showpacs]{revtex4}
\usepackage{graphicx}% Include figure files
\usepackage{dcolumn}% Align table columns on decimal point
\usepackage{bm}% bold math

\begin{document}

%\preprint{}

\title{Spectroscopy of Rb$_{2}$ dimers in solid $^{4}$He}



\author{P. Moroshkin}
\email[]{peter.moroshkin@unifr.ch}
\author{A. Hofer}
\author{S. Ulzega}
\author{A. Weis}

\affiliation{D\'{e}partement de Physique, Universit\'{e} de Fribourg,
Chemin du Mus\'{e}e 3, 1700 Fribourg, Switzerland}
\homepage[]{www.unifr.ch/physics/frap/}

\date{\today}

\begin{abstract}
We present experimental and theoretical studies of the absorption, emission and photodissociation spectra of
Rb$_{2}$ molecules in solid helium. We have identified 11 absorption bands of Rb$_{2}$. All laser-excited
molecular states are quenched by the interaction with the He matrix. The quenching results in efficient
population of a metastable (1)$^{3}\Pi_{u}$ state, which emits fluorescence at 1042 nm. In order to explain the
fluorescence at the forbidden transition and its time dependence we propose a new molecular exciplex
Rb$_{2}(^{3}\Pi_{u})$He$_{2}$. We have also found evidence for the formation of diatomic bubble states following
photodissociation of Rb$_{2}$.
\end{abstract}

\pacs{33.20.-t, 33.50.-j, 33.80.Gj, 67.80.Mg, 67.80.-s}

\maketitle


\section{Introduction\label{seq:introduction}}

The spectroscopy of alkali dimers, in particular of the Rb$_{2}$ molecule is a well established area of research
 \cite{Drummond;SpectraKineticsRb2,Pichler;InterferenceDiffuseContinuaRb2,Amiot;LaserInducedFluorescenceRb2,Lee;DirectObservation2trpPiRb2}. Recently it has attracted
renewed attention in connection with the creation of ultracold
molecules via the photoassociation of ultracold atoms in
magneto-optical traps
\cite{Fioretti;ColdRbMoleculeFormation,Gutterres;DeterminationRb5PDipoleMatrixElement}.
In such experiments the ultracold molecules are usually produced
in their lowest triplet state ($a^{3}\Sigma$), and their transfer
into the zero-vibration singlet ground state $X^{1}\Sigma$ remains
a major problem
\cite{DeMille;ProspectsProductionXRbSc,Stwalley;EfficientConversionFeshbach,Azizi;ProspectsFormationUltracold}.
Another recently developed technique
\cite{Scoles;AlkaliDimersTripletManifold,Bunermann;SpectroscopyCsAttachedNanodroplets,Ernst;TripletStatesRbDimers,Reho;Dynamics3PigStateK2}
involves the attachment of several alkali atoms to a superfluid
helium nano-droplet, where they cool down to 0.38 K and may form a
dimer. The energy released in the formation of a strongly bound
molecule in its singlet ground state usually results in the
complete evaporation of the nano-droplet, therefore only weakly
bound triplet states could be observed in most of these studies.
There have been a few reports on alkali dimers in cryogenic
rare-gas matrices. Takahashi \textit{et. al.}
\cite{Yabuzaki;SpectroscopyAlkaliAtomsAndMolecules} observed
emission of Na$_{2}$ molecules in superfluid helium, and some
absorption bands of Li$_{2}$ and Na$_{2}$ in solid xenon were
reported in \cite{Welker;OpticalAbsorptionMatrixIsolatedLiNaAg}.
Theoretical studies were reported for Na$_{2}$ in solid argon
\cite{Gross;TheoreticalStudyNaArgonMatrix,Gervais;NaClustersinRareGasMatrix}.

In our earlier publications, the spectroscopy of Cs and Rb atoms in solid helium has been well established
\cite{Kanorsky;PressureshiftHelium,Moroshkin;ExciplexesJCP,Eichler;OptDetecNonradAlkaliInSolHe,Hofer;RbHeExciplex}.
The absorption of laser radiation at the wavelengths of the Cs $D_{1}$ and $D_{2}$ transitions, blueshifted by
the interaction with the He matrix, induces fluorescence on the atomic D$_{1}$ line and emission bands from
bound-free transitions in Cs$^{\ast}$He$_{2}$ and Cs$^{\ast}$He$_{7}$ exciplexes. Fluorescence from Rb atoms in
solid helium is quenched by the interaction with the matrix \cite{Yabuzaki;Quenching}. The quenching is
attributed to the very fast formation of Rb$^{\ast}$He$_{N}$ exciplexes.  A weak absorption on the Rb D$_{1}$
transition at 760 nm has been detected in a transmission experiment \cite{Eichler;OptDetecNonradAlkaliInSolHe}.
Only recently \cite{Hofer;RbHeExciplex} a weak fluorescence from the Rb $D_{1}$ and $D_{2}$ transitions was
obtained in solid He, together with the emission of Rb$^{\ast}$He$_{N}$ exciplexes. No fluorescence from nor
absorption by Rb$_2$ molecules in bulk solid or liquid He has been reported so far. Here we present the first
experimental study of Rb$_2$ dimers in solid helium.

\section{Experiment\label{seq:experiment}}

\subsection{Experimental setup}

\begin{figure}[tbp]
\includegraphics[width=8.5cm]{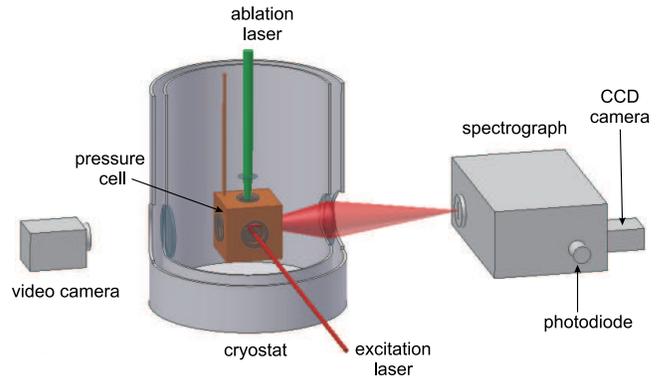}
\caption{Vertical section through the helium bath cryostat showing the pressure cell. A pulsed focused laser beam
(532 nm) from the top is used for the implantation, and a non-focused horizontal laser beam is used for the
excitation of the implanted atoms/molecules. The video camera is used to monitor crystal growth and the
implantation process.} \label{fig:implantation}
\end{figure}

The experimental setup is illustrated in
Fig.~\ref{fig:implantation}. A solid $^{4}$He matrix doped with Rb
atoms was produced by the technique described in our earlier
papers \cite
{Kanorsky;OpticalSpectroscopySolidHelium,Moroshkin;ExciplexesJCP}.
The measurements were performed in the hcp and bcc crystalline
phases of solid $^{4}$He in the pressure range of $27-36$~bar, at
a temperature of $1.5$~K . The sample was produced in a helium
pressure cell immersed in superfluid helium cooled by pumping on
the helium bath. Windows provide optical access from three
orthogonal directions. A helium crystal is grown inside the
pressure cell, by condensing and then solidifying pressurized
helium gas from an external reservoir. The helium host matrix is
then doped with rubidium atoms, as shown in
Fig.~\ref{fig:implantation}, by means of laser ablation with the
second harmonic of a pulsed frequency-doubled Nd:YAG-laser beam
(or the signal beam of the OPO described below) focused onto a
solid target containing pure natural Rb or a Rb-Cs mixture.

We have studied the absorption spectrum of the sample in the range of 400-1000 nm. In order to cover such a broad
range, three sources of coherent radiation were used: (i) the signal beam from an optical parametric oscillator
(OPO, OPTA GmbH model BBO-355-vis/IR), pumped by the third harmonic (354~nm) of a pulsed Nd:YAG laser for 400-700
nm; (ii) the idler beam of the same OPO for 780-1000 nm, and (iii) a tunable cw Ti:Sa laser pumped by a
Nd:YVO$_{3}$ laser for 720-780 nm. The excitation laser beams (ii) and (iii) cross the pressure cell in a
horizontal direction, passing through the volume of maximal doping. The beam (i) illuminates the sample from the
top of the cryostat. In this configuration the signal beam of the OPO (pulse energy of 2 mJ, pulse width of 5 ns,
wavelength in the range of 500-670 nm) can also be used as implantation laser.

Fluorescence light from the sample volume (approx. 3~mm$^{3}$) is collected by a lens located in the cryostat and
then analyzed by a grating spectrograph (Oriel, model MS257) with a resolution of 2 nm equipped with a CCD
camera. In addition to spectral measurements we also performed  time-resolved studies of the molecular
fluorescence, for which we used a Si photodiode mounted after a second output slit of the spectrograph. The
time-resolution of this system is limited by the 100 kHz bandwidth of the preamplifier.

\subsection{Observed emission spectra}
\begin{figure}[tbp]
\includegraphics[width=8.5cm]{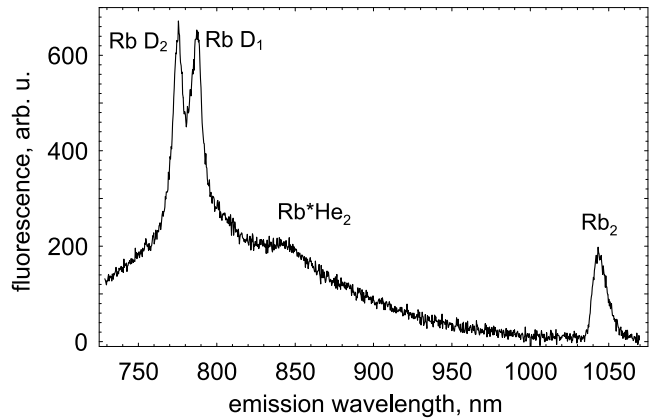}
\caption{Measured emission spectrum from Rb implanted in solid helium (T=1.5 K, p=31 bar). Excitation by the focused signal beam of the OPO at
570 nm.} \label{fig:emissionspectrum}
\end{figure}

The fluorescence spectrum observed in experiments with a crystal doped from a solid Rb target contains features
which can be associated with atomic Rb, with Rb$_2$ dimers and with Rb$^{\ast}$He$_{N}$ exciplexes
(Fig.~\ref{fig:emissionspectrum}). Recently we have done a detailed spectroscopic study of the Rb-He exciplexes
\cite{Hofer;RbHeExciplex}. We restrict the present discussion to Rb and Rb$_{2}$. A typical emission spectrum
obtained under excitation at 570 nm is shown in Fig.~\ref{fig:emissionspectrum}, in which one distinguishes three
spectral features. The two components of the narrow doublet at 780 nm are easily identified as the $D_{1}$ and
$D_{2}$ emission lines of atomic Rb. They are separated by 14 nm, which corresponds to the fine-structure
splitting of the Rb 5P$_J$ states in solid helium. Both components of the doublet are strongly broadened and
blueshifted with respect to the free Rb atom due to the interaction with the surrounding He. Both the shift and
the spectral width depend on the He pressure. The intensity of these two lines is orders of magnitude lower than
the intensity of the $D_{1}$ line of Cs recorded under similar conditions \cite{Moroshkin;ExciplexesJCP}. In a
previous study \cite{Eichler;OptDetecNonradAlkaliInSolHe} the $D_{1}$ line of Rb was excited by a cw Ti:Sa laser
in an absorption experiment at 760 nm, but fluorescence was overlooked in that experiment, probably because of
the strong intensity of scattered laser radiation in this part of the spectrum. We attribute the central spectral
feature of Fig.~\ref{fig:emissionspectrum} to Rb*He$_{2}$ exciplexes, formed by the atomic 5$P$ states as
discussed in detail elsewhere \cite{Hofer;RbHeExciplex}. The rightmost spectral feature at 1042 nm originates
from the Rb$_2$ dimer. It is the only molecular band that could be observed in the fluorescence spectrum in the
range of 500-1600 nm, when exciting the sample in the range of 450-900 nm.

\subsection{Excitation spectra}

\begin{figure}[tbp]
\includegraphics[width=8.5cm]{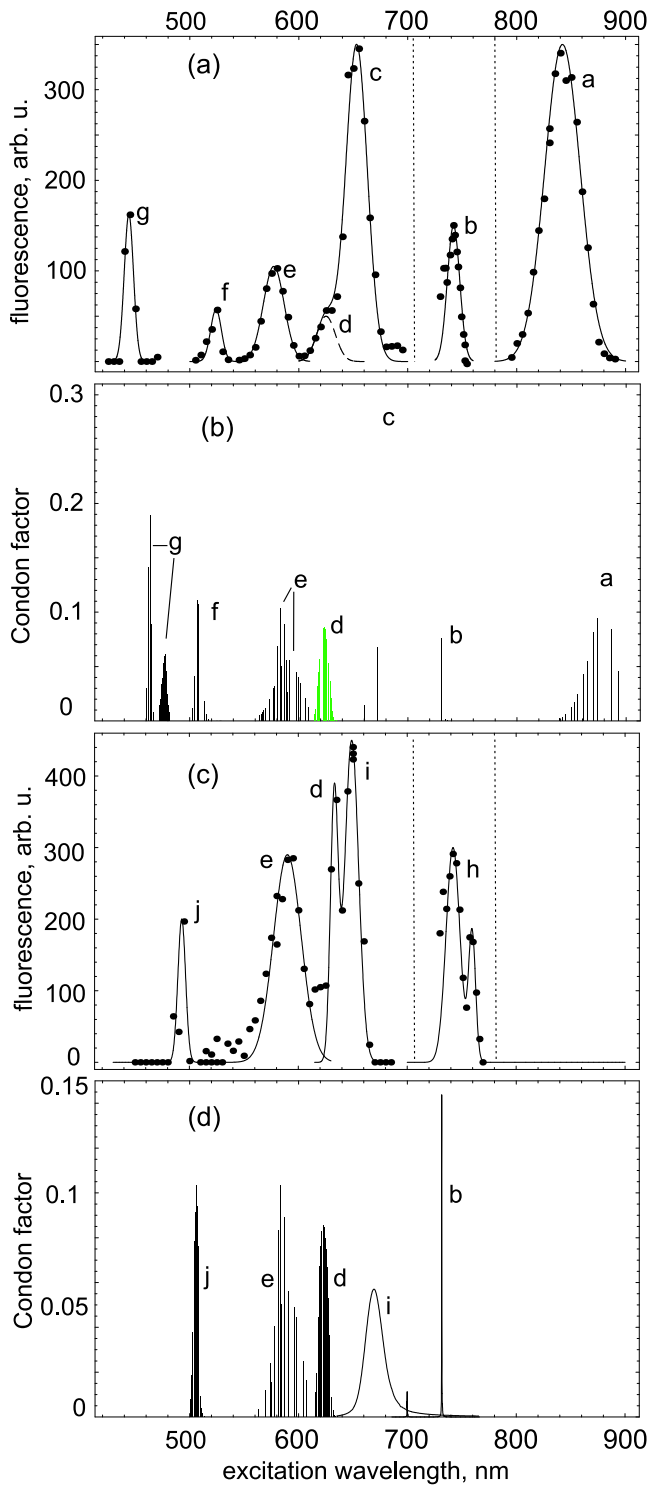}
\caption{(a) measured excitation spectrum of the molecular fluorescence at 1042 nm, (c) measured
photodissociation spectrum of the Rb dimer (detected via the atomic Rb fluorescence at 780 nm). (b), (d) are
calculated Condon factors for the transitions from the $X^{1}\Sigma_{g}$ and (1)$^{3}\Sigma_{u}$ states
populating the (1)$^{3}\Pi_{u}$ state (b) and the atomic 5$P_{1/2}$ state (d). Experimental data are shown as
points, and the solid lines are fitted Gaussians. The assignment of all peaks is given in
Table~\ref{table:shifts} and in the text of Sec.~\ref{seq:theory}. The dotted vertical lines mark the spectral
regions covered by the 3 excitation sources described in the text. The vertical scales differ for the different
intervals.} \label{fig:excitationspec}
\end{figure}

The molecular emission at 1042 nm can be excited on a number of resonant absorption bands. The excitation
spectrum of the 1042 nm emission is shown in Fig.~\ref{fig:excitationspec}(a). The rightmost band centered at 842
nm (feature "a") is the strongest band observed in the present experiments. The band at 740 nm (labelled "b")
overlaps with the D$_{2}$ absorption line of atomic Rb. As a rule, the signal is much larger when the beam is
focused and illuminates the sample in the vertical direction. However, some of the other bands with maxima at
650, 580, 525, and 445 nm could even be observed with a non-focused beam. In addition, the blue wing of the band
at 650 nm has a shoulder (labelled "d"), which can be interpreted as a weaker band partially overlapping with the
650 nm band. At higher helium pressures this band shifts towards the blue at a faster rate than the stronger
band. At 35 bar its maximum is at 625 nm and the two bands become resolved.

The fact that the emission peak at 1042 nm appears under excitation at very different wavelengths and that no
other molecular emission is observed suggests that all molecular excited states are quenched by the interaction
with the helium matrix. The excitation of any state seems to be followed by very fast radiationless or far
infrared transitions to a low-lying state which decays by emitting 1042 nm light.

Fig.~\ref{fig:excitationspec}(c) shows the excitation spectrum for atomic emission on the Rb $D_{1}/D_{2}$
doublet (actually the detection spectrograph was set to 780 nm, which corresponds to $D_1$ emission). The
features "j", "e", "d", and "i" show that atomic emission can not only be excited by absorption at the atomic
resonance lines (740 and 760 nm respectively), but also at 650, 580, and 495 nm. We assign these bands to
absorption by molecular state(s) of Rb$_{2}$ which dissociate into one ground-state and one excited state atom.

\subsection{Time resolved fluorescence}

\begin{figure}[tbp]
\includegraphics[width=7.5 cm]{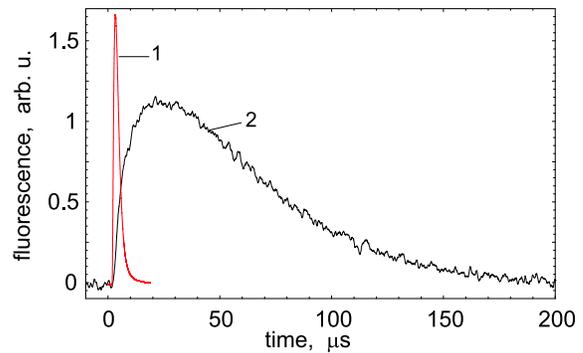}
\caption{Measured pulse-shapes of the molecular fluorescence at 1042 nm (curve 2) and scattered of laser light at
842 nm (curve 1). The actual width of the laser pulse (5 ns) is not resolved so that curve~1 represents the time
resolution of the photodetector.}\label{fig:pulseshape}
\end{figure}

We measured the time dependence of  the molecular fluorescence following pulsed excitation at selected excitation
and emission wavelengths. A characteristic fluorescence pulse shape is shown in Fig.~\ref{fig:pulseshape}. The
atomic and exciplex emission decay-times are too short to be resolved by our detection system. Both channels
yield a fluorescence pulse which coincides with that of the scattered laser light (curve 1 in
Fig.~\ref{fig:pulseshape}) and thus reflects the time resolution (FWHM = 3 $\mu$s) of the detector. This is
consistent with other studies of atomic Rb and Cs in liquid He \cite{Yabuzaki;Quenching}, and our own
(unpublished) results on Cs in solid He in which atomic lifetimes on the order of a few 10 ns were observed. The
molecular fluorescence at 1042 nm shows an exponential decay with a characteristic time of 45 $\mu$s that is well
resolved (curve 2 in Fig.~\ref{fig:pulseshape}). The fluorescence pulse shows also an intriguing finite rise-time
of approximately 15 $\mu$s. Pulse shapes with identical decay and rise-times could be measured when exciting
either at 840 or at 650 nm.

\subsection{Emission spectra at the phase transition}

We have performed experiments at different helium pressures in the range of 26-36 bar. At 1.5 K, two phase
transitions occur within this range \cite{Vignos;NewSolidPhaseHe}. The solidification of superfluid He at 26.4
bar produces first a body-centered cubic (bcc) crystalline structure, and then a hexagonal close-packed (hcp)
structure at 26.8 bar. The standard implantation procedure described above produces the best results, i.e., the
largest density of implanted species and hence of spectroscopic signals when applied at 29.5 bar in the hcp
crystal. By releasing the pressure after the doping process to below 27 bar we were able to observe molecular and
atomic fluorescence in the bcc phase and even during the transition to the liquid phase.

\begin{figure}[tbp]
\includegraphics[width=7.5cm]{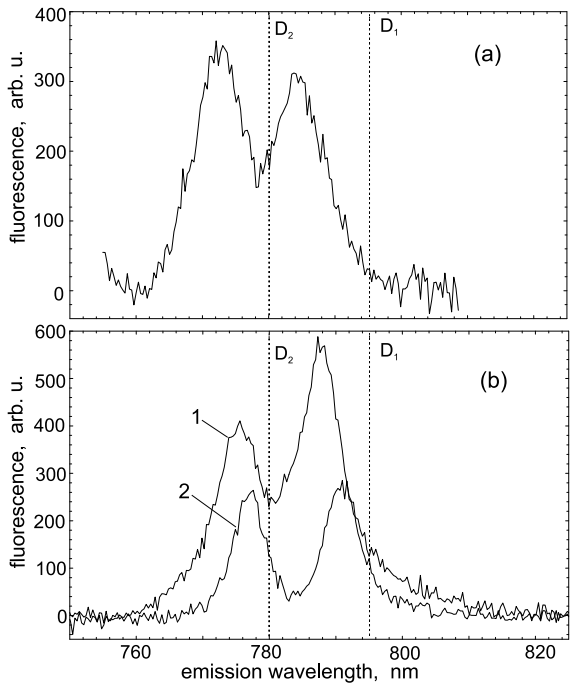}
\caption{Measured emission spectrum of atomic Rb produced by excitation of the atomic 5$P_{3/2}$ state at 740
nm~(a) and by photodissociation of Rb$_{2}$ following excitation at 650 nm (b): curve 1 - hcp phase at $p$ = 30
bar, curve 2 - during the solid-liquid phase transition at $p$ = 26.4 bar. Vertical dashed lines mark the
wavelengths of the free atomic $D_{1}$ and $D_{2}$ lines at 795 and 780 nm.} \label{fig:atomemis}
\end{figure}

When atomic Rb in a hcp crystal at 30 bar is excited at 740 nm  the $D_{1}$/$D_{2}$ fluorescence spectrum
(Fig.~\ref{fig:atomemis}(a)) is blue-shifted by approximately 10 nm and strongly broadened with respect to the
free atomic lines. When the same fluorescing transitions are observed after photodissociation of Rb$_{2}$ at 650
nm under otherwise identical conditions the emission lines shown as curve 1 in Fig.~\ref{fig:atomemis}(b) are
observed. The spectrum shows a reduced blue shift of approximately 7 nm. These fluorescence spectra  exhibit
interesting features, when recorded during the melting process. The blueshift and the intensity of the
fluorescence following absorption on the photodissociating band at 650 nm decrease with decreasing He pressure,
similar to the behavior observed on the atomic fluorescence of Cs (excited on the atomic $D_1$ line) in solid He
\cite{Moroshkin;ExciplexesJCP}. However, unlike the observation of the purely atomic process in cesium-doped
helium, the intensity rises again when the (photodissociation-induced) fluorescence is recorded during the phase
transition to the liquid. The fluorescence recorded under the latter conditions is shown as curve~2 in
Fig.~\ref{fig:atomemis}(b). One sees that the spectral width of each component of the atomic fluorescence doublet
(curve 2) coming from that structure is smaller and that their intensity ratio differs from the one observed in
hcp helium (curve1). The blueshift of both spectral lines under those conditions is only 4.0\,nm, much smaller
than in solid He, but also significantly smaller than in pressurized (25 bar) superfluid He
\cite{Kinoshita;OpitcalPropAlkaliAtomsPressSuperfluidHe}, when excited at the atomic $D_1$ line (760 nm). We
could not observe any fluorescence after the complete melting of the crystal.

The molecular fluorescence band at 1040\,nm (excited at 650\,nm) exhibits a qualitatively different behavior at
the same phase transition. The spectral position and the width of the band do not show a measurable pressure
dependence and the intensity quickly vanishes at the transition from the bcc phase to the liquid. We could not
detect any molecular emission induced by excitation at 650 nm, under the same conditions in which curve 2 of
Fig.~\ref{fig:atomemis}(b) was obtained.

We have performed the same measurements when the sample is irradiated with 580\,nm light. Here the behavior is
again quite different. The atomic fluorescence disappears already at the beginning of the bcc to liquid phase
transition, while, at the same time, the molecular fluorescence can be observed until the complete melting. This
interesting phenomenon will be discussed below.

\section{Theoretical model\label{seq:theory}}

The dimer Rb$_{2}$ is the best candidate for describing most of
the observed spectral lines described above, and we will
substantiate this assignment by discussing its absorption and
fluorescence spectra in detail.

\subsection{Molecular potentials}

\begin{figure}[tbp]
\includegraphics[width=8.5cm]{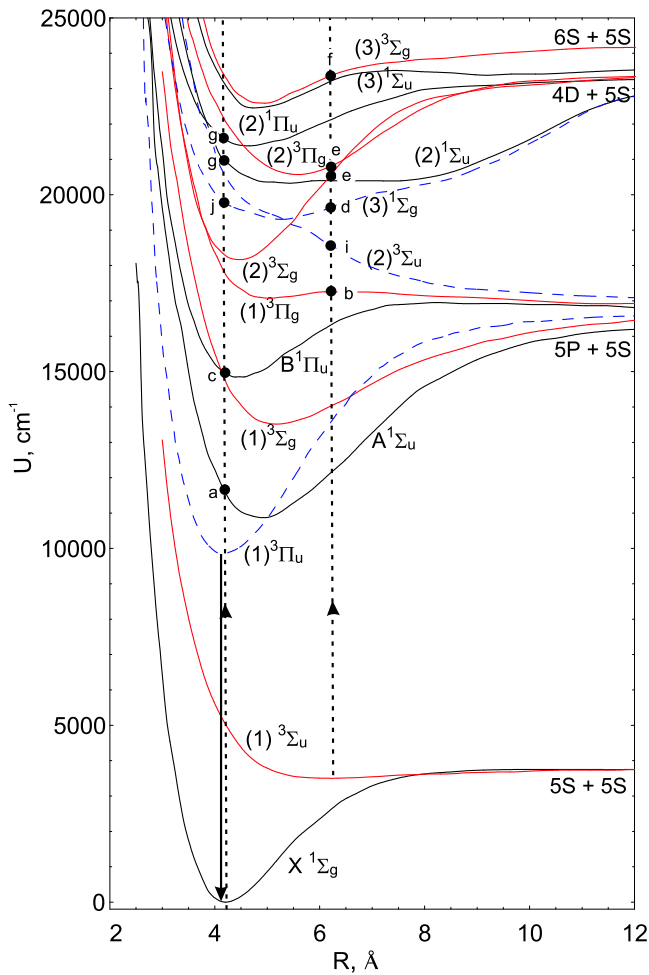}
\caption{Theoretical \textit{ab initio} potential curves of the Rb$_{2}$ molecule, from
\cite{Park;TheoreticalRb2Potentials}. Solid lines - states connected by the allowed transitions to the one of two
ground-states; dashed lines - states without the allowed transition to the ground-state (only the states
discussed in this work are shown). Two vertical dashed lines mark the vertical transitions from the singlet and
triplet ground-state at 4.2 and 6.2 \AA. and the vertical solid arrow indicates the emission from the metastable
state.} \label{fig:potentials}
\end{figure}

Theoretical \textit{ab initio} potential curves of the Rb$_{2}$ molecule are available in the literature
\cite{Spiegelmann;TheoreticalRb2Potentials,Park;TheoreticalRb2Potentials}. The potentials from
\cite{Park;TheoreticalRb2Potentials} are shown in Fig.~\ref{fig:potentials}. The ground state of Rb$_2$ has two
components: a deeply bound (covalent) singlet state $X^{1}\Sigma_{g}$ and a weakly bound (van der Waals) triplet
state $(1) ^{3}\Sigma_{u}$.

The wavefunctions of the molecule's oscillations depend on a
single parameter $r$, the internuclear separation in the diatomic
molecule, and can be found by solving the one-dimensional
Schr\"{o}dinger equation
\begin{equation}  \label{eq:Schroedinger}
-\frac{\hbar^{2}}{2m} \frac{d^{2}\Psi_\alpha^{(v)}}{dr^{2}}+V_\alpha(r)\Psi_\alpha^{(v)}=\omega_{v}\Psi_\alpha^{(v)},
\end{equation}
for the corresponding potential $V_\alpha(r)$, where $\omega_{v}$ is the eigenenergy of the vibrational state $v$.

The transition probabilities between vibrational states of a given electronic transition are given by the
Franck-Condon factors, i.e., the overlap integrals between the wavefunctions of excited ($\Psi_{e}$) and ground
($\Psi_{g}$) states. There is an important simplification resulting from the fact that at the temperature of our
experiments (1.5 K) only the lowest molecular vibration states are populated. Therefore the absorption spectrum
consists only of transitions from the $v=0$ vibration level in the electronic ground state to all vibrational
levels $v'$ in the excited state, for which the Franck-Condon factors are given by

\begin{equation}  \label{eq:FranckCondonexc}
F_{exc}^{(v')}=|\int \Psi_{g}^{(0)}(r)\Psi_{e}^{(v')}(r)dr|^{2},
v'=0, 1, 2, ...
\end{equation}

In the same way, the emission spectrum consists of transitions
from the $v'=0$ vibration level in the electronic excited state to
all vibrational levels $v$ in the ground state.

\begin{equation}  \label{eq:FranckCondonemis}
F_{emis}^{(v)}=|\int \Psi_{g}^{(v)}(r)\Psi_{e}^{(0)}(r)dr|^{2}, v=0, 1, 2, ...
\end{equation}

When calculating the Franck-Condon factors one has to take into account the selection rules, which allow only
transitions between terms of the same multiplicity, while transitions between singlet and triplet states are
forbidden. In most spectroscopic experiments \cite{Amiot;LaserInducedFluorescenceRb2}, the alkali dimers are
produced in a hot and dense atomic vapor in which only the deeply bound $X^{1}\Sigma_{g}$ state is populated.
Therefore only the transitions between singlet states have been studied in detail and the corresponding potential
curves evaluated. As for the triplet states, the available potential curves are obtained mainly from \textit{ab
initio} calculations.

\subsection{Emission spectrum}

The long decay-time of the molecular fluorescence suggests that it originates from a metastable state and that
the corresponding transition is forbidden in the free molecule. Inspection of the potential energy diagram in
Fig.~\ref{fig:potentials} reveals the (1)$^{3}\Pi_{u}$ state to be a good candidate for the excited state of the
fluorescing transition. It is the lowest lying excited state of the molecule, and it can be populated by a
cascade of radiationless transitions following laser excitation to any other higher lying state. In the free
molecule the transition to the singlet ground-state $X ^{1}\Sigma_{g}$ is forbidden by multiplicity and the
triplet-triplet transition to the (1)$^{3}\Sigma_{u}$ state by parity. The absence of fluorescence from other
excited states suggests that in solid He these states are quickly quenched and that all excited population is
collected in the long-lived (1)$^{3}\Pi_{u}$ state, which finally decays towards the $X ^{1}\Sigma_{g}$
ground-state. A similar effect was reported in spectroscopic studies of Hg$_{2}$ dimers in a cryogenic Ar matrix
\cite{Chergui;SpectroscopyRelaxationHgNeArXe}, where only the emission from the metastable A0$_{g}^{+}$ state was
observed under excitation on four different allowed transitions.

\begin{figure}[tbp]
\includegraphics[width=8. cm]{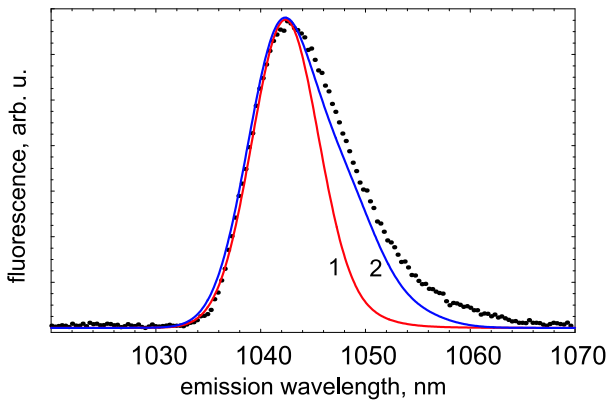}
\caption{Experimental (points) and calculated (solid curves) emission spectra of (1)$^{3}\Pi_{u} \rightarrow X
^{1}\Sigma_{g}$ transition in Rb$_{2}$. Curve 1 was obtained by shifting the calculated curve by 20 nm to the
red. Curve 2 is obtained by shifting the excited state potential curve by 0.12 \AA~ towards larger interatomic
distances, and then shifting again the calculated spectrum by 20 nm to the red.} \label{fig:fitemiss}
\end{figure}

The potentials in Fig.~\ref{fig:potentials} show that the minimum of the (1)$^{3}\Pi_{u}$ potential lies almost
at the same internuclear distance as the minimum of the ground state. The overlap integral between the lowest
vibrational states $v'=v=0$ is therefore exceptionally large and quickly vanishes with increasing $v'$. In fact
only 0$\rightarrow$0 and 0$\rightarrow$1 transitions produce non-negligible contributions to the spectrum. In
order to model the observed emission band, we represent these two components by Gaussian curves with a spectral
width (FWHM) of 7.5 nm. This width was chosen to match the left side of the experimental spectrum as shown in
Fig.~\ref{fig:fitemiss} and is typical for atomic emission lines in solid helium. The amplitudes of the curves
are given by the calculated Franck-Condon factors.

The calculated emission band has its peak at 1022 nm, whereas the experimental band peaks at 1042 nm. This shift
of 20 nm can be attributed either to the interaction of the molecule with the helium matrix, or to the
imprecision of the theoretical molecular potentials. Atomic emission lines in solid and liquid He are known to be
shifted by 10-15 nm towards shorter wavelengths, and to show an increase of this shift with the increase of He
pressure. However, our observations of the molecular fluorescence band discussed here show that its spectral
position and its lineshape do not depend on pressure, which leads us to conclude that the observed wavelength is
close to the one of the free molecule. As the ground-state potential curve is very well known from theory and
experiments \cite{Amiot;LaserInducedFluorescenceRb2XSigmaG,Amiot;LaserInducedFluorescenceRb2}, one therefore has
to suspect imprecisions of the (1)$^{3}\Pi_{u}$ potential curve to be responsible for the 20 nm shift. There are
no detailed spectroscopic studies of this state. Only the mutual perturbation of the $A ^{1}\Sigma_{u}$ and
(1)$^{3}\Pi_{u}$ states in the vicinity of their crossing point at 5.1~\AA~was studied
\cite{Zhang;QuantumWavePacketDynamicsRb2}. The observed fluorescence spectrum (dots) and the calculated band
shifted by 20 nm (curve 1), are compared in Fig.~\ref{fig:fitemiss}, after scaling the amplitude of the
calculated spectrum. The 20 nm shift suggests that the depth of the calculated potential well is underestimated
by 190 cm$^{-1}$.

The observed emission spectrum has a pronounced asymmetry. It may result from the overlap of several (two)
vibronic transitions, as discussed above. However, when each component is modelled by a symmetric Gaussian curve,
the resulting asymmetry is smaller than that of the experimental spectrum (see Fig.~\ref{fig:fitemiss}). This
discrepancy can be reduced if one shifts the minimum of the excited state potential with respect to that of the
ground state. Curve 2 in Fig.~\ref{fig:fitemiss} was obtained by shifting this potential curve by 0.12 \AA~
towards larger interatomic distances and yields a better agreement with the experimental data. The alternative
interpretation of the discrepancy between the measured and calculated emission spectra is presented below, in
Sec.~\ref{seq:MolecularExciplex}.

\subsection{Molecular absorption spectrum}

When comparing the calculated Franck-Condon factors of the absorption bands shown in
Fig.~\ref{fig:excitationspec}(b) to the experimentally measured spectra of Fig.~\ref{fig:excitationspec}(a) one
can identify all observed bands. The calculated wavelengths of the band maxima are presented in the third column
of Table~\ref{table:shifts}. The relative intensities can not be predicted by our model which does not consider
the transition dipole moments. We have fitted the experimental spectra with Gaussian curves. The results of the
fits are shown in Fig.~\ref{fig:excitationspec}(a) as solid lines and the positions of the band maxima are
collected in the fourth column of Table~\ref{table:shifts}.

\begin{table}
  \centering
\begin{tabular}{c||ccccc}
band& &label&$\lambda_{theor}$&$\lambda_{exper}$\\\hline\hline $X ^{1}\Sigma_{g}\rightarrow A ^{1}\Sigma_{u}$&
&a&878&842\\\hline (1)$^{3}\Sigma_{u} \rightarrow$ (1)$^{3}\Pi_{g}$& &b&735&742\\\hline $X^{1}\Sigma_{g}
\rightarrow B^{1}\Pi_{u}$& &c&664&653\\\hline (1)$^{3}\Sigma_{u} \rightarrow$ (3)$^{1}\Sigma_{g}$&
&d&623&622\\\hline (1)$^{3}\Sigma_{u} \rightarrow$ (2)$^{3}\Sigma_{g}$& &e&590&580\\\hline (1)$^{3}\Sigma_{u}
\rightarrow$ (2)$^{3}\Pi_{g}$& &e&586&580\\\hline (1)$^{3}\Sigma_{u} \rightarrow$ (3)$^{3}\Sigma_{g}$&
&f&507&524\\\hline $X^{1}\Sigma_{g} \rightarrow$ (2)$^{1}\Sigma_{u}$& &g&477&445\\\hline $X^{1}\Sigma_{g}
\rightarrow$ (2)$^{1}\Pi_{u}$& &g&464&445\\\hline (1)$^{3}\Sigma_{u} \rightarrow$ (2)$^{3}\Sigma_{u}$&
&i&670&649\\\hline $X^{1}\Sigma_{g} \rightarrow$ (3)$^{1}\Sigma_{g}$& &j&505&493\\\hline \label{table:shifts}
\end{tabular}
\caption{Calculated and measured wavelengths of Rb$_{2}$ absorption bands (in nm). The second column shows the
labels of the corresponding peaks in Fig.~\ref{fig:excitationspec}.}
\end{table}

The strongest absorption band at 842 nm (band "a" in Fig.~\ref{fig:excitationspec}(a)) belongs to the $X
^{1}\Sigma_{g}\rightarrow A ^{1}\Sigma_{u}$ transition. The experimentally measured absorption band is
blueshifted with respect to the calculated one by 35~nm. This shift can again be attributed either to the
influence of the He-matrix, or to an uncertainty in the excited state potential curve. The band at 740 nm (band
"b" in Fig.~\ref{fig:excitationspec}(a)) is assigned to the (1)$^{3}\Sigma_{u} \rightarrow$ (1)$^{3}\Pi_{g}$
transition. Both the excited and the ground state have very shallow potential wells and therefore the transition
wavelength is very close to that of the atomic D$_{2}$ absorption line (the red component of the doublet "h" in
Fig.~\ref{fig:excitationspec}(c)).

Another strong absorption band is observed at 653 nm (band "c" in Fig.~\ref{fig:excitationspec}(a)). We identify
it as the $X^{1}\Sigma_{g} \rightarrow B^{1}\Pi_{u}$ band. The calculated band position is at 664 nm. The
potential curves of the  involved states are well known from experiments
\cite{Amiot;LaserInducedFluorescenceRb2XSigmaG,Amiot;OpticalDoubleResonanceRb2BState}. The blueshift due to the
interaction with the He-matrix is 11 nm, significantly smaller than that of the atomic lines and of the $X
^{1}\Sigma_{g}\rightarrow A ^{1}\Sigma_{u}$ band.

The weak absorption band labelled "d" in Fig.~\ref{fig:excitationspec}(a) lies in a spectral region in which no
allowed molecular transitions exist. Its wavelength strongly depends on the helium pressure. We suppose that this
band (at 622 nm) is due to the forbidden transition (1)$^{3}\Sigma_{u} \rightarrow$ (3)$^{1}\Sigma_{g}$ from the
triplet ground-state.

The observed absorption band at 580 nm (peak "e" in Fig.~\ref{fig:excitationspec}) is a superposition of two
allowed triplet-triplet bands: (1)$^{3}\Sigma_{u} \rightarrow$ (2)$^{3}\Sigma_{g}$ and (1)$^{3}\Sigma_{u}
\rightarrow$ (2)$^{3}\Pi_{g}$. None of these states has been studied experimentally before. The calculated
wavelengths of both transitions are longer than the experimental one by 6 and 10 nm respectively.

The absorption band at 525 nm (peak "f" in Fig.~\ref{fig:excitationspec}(a)) can be attributed to another
triplet-triplet system: (1)$^{3}\Sigma_{u} \rightarrow$ (3)$^{3}\Sigma_{g}$. The calculated position of this band
is at 507 nm, \textit{i. e.} the interaction with the matrix shifts it to the red, unlike all other observed
bands.

The narrow band at 445 nm (peak "g" in Fig.~\ref{fig:excitationspec}(a)) can be assigned to allowed transitions
from the singlet ground state. The $X^{1}\Sigma_{g} \rightarrow$ (2)$^{1}\Sigma_{u}$ and $X^{1}\Sigma_{g}
\rightarrow$ (2)$^{1}\Pi_{u}$ bands with calculated wavelengths of 477 and 465 nm respectively are good
candidates to explain this band.

\subsection{Speculation about the formation of a molecular  Rb$_{2}^{\ast}$He$_{N}$ exciplex\label{seq:MolecularExciplex}}

We have shown above that every observed excitation of the Rb$_2$ dimer finally leads to a population of the
(1)$^{3}\Pi_{u}$ state. The laser-excited states do not fluoresce at the excitation wavelength, nor do they emit
fluorescence in the near (red or blue) vicinity of the excitation wavelength as atoms, e. g., typically do. The
quenching of all upper states thus seems to occur on a time scale which is much shorter than their radiative
lifetimes. This deexcitation proceeds via a direct or a cascade-like radiationless transition or via transitions
emitting far-infrared light, which is not accessible by our spectrometer. Only the final metastable state
(1)$^{3}\Pi_{u}$, perturbed by the interaction with the surrounding He, lives long enough to emit fluorescence
with a lifetime of 45 $\mu$s.

Besides this long lifeteime we have observed an intriguingly long (15 $\mu$s) rise-time of the fluorescence
signal which indicates that the fluorescing state is populated via some intermediate metastable level.  This
state must lie below the $A^{1}\Sigma_{u}$ state, the lowest lying level after whose excitation one observes the
1042 nm fluorescence. In Fig.~\ref{fig:potentials} one sees that no such state exists in the free Rb$_{2}$
molecule.

\begin{figure}[tbp]
\includegraphics[width=6. cm]{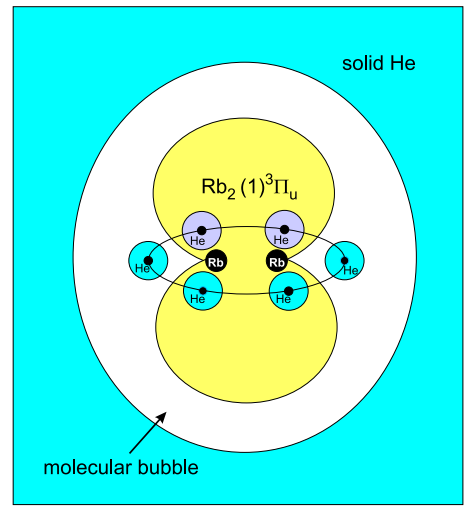}
\caption{Structure of the molecular Rb$_{2}^{\ast}$He$_{N}$ exciplex in which several He atoms from the interface
of the molecular bubble produce a bound or quasi-bound state with the rotationally quenched metastable Rb$_{2}$
((1)$^{3}\Pi_{u}$) molecule. This structure is proposed for explaining the puzzling time-dependence of the
molecular fluorescence.} \label{fig:molexciplex}
\end{figure}

For this reason we tentatively explain the observation in terms of a relatively slow conformational change of the
system formed by the metastable Rb$_{2}$($^{3}\Pi_{u}$) molecule and the surrounding helium atoms (molecular
bubble). The quantitative modelling of the interaction between the defect molecule and the He crystal is beyond
the scope of the present paper. However, such a theoretical study was reported recently
\cite{Eloranta;SolvationTripletRydbergH2} for the hydrogen molecule embedded in liquid He. For the metastable
$c^{3}\Pi_{u}$ state which is analogous to the (1)$^{3}\Pi_{u}$ state of Rb$_{2}$ the authors of
\cite{Eloranta;SolvationTripletRydbergH2} suggest an exciplex structure consisting of several He atoms bound
around the waist of the dumbbell-shaped molecular orbital of H$_{2}$. This structure closely resembles that of
Cs(6$P_{1/2}$)He$_{7}$ and Rb(5$P_{1/2}$)He$_{6}$ exciplexes observed in our recent experiments
\cite{Moroshkin;ExciplexesJCP,Hofer;RbHeExciplex} in solid He, it is shown schematically in
Fig.~\ref{fig:molexciplex}. Here we suggest that a similar molecular exciplex Rb$_{2}^{\ast}$He$_{N}$ could be
formed by the (1)$^{3}\Pi_{u}$ state of the Rb$_{2}$ molecule in solid He.

The metastable state of the dimer is prepared in a relatively large molecular bubble. In order to leave the
bubble surface and to move towards the dimer, the He atoms have to tunnel through a potential barrier. The
tunnelling time may be as long as a few microseconds, which would explain the slow onset of the fluorescence. The
perturbation of the dimer by the bound He atoms lifts the selection rule that forbids the radiative transition to
the singlet ground-state. Unlike the atomic exciplexes studied in
\cite{Moroshkin;ExciplexesJCP,Hofer;RbHeExciplex}, the suggested molecular exciplex has a very long life-time and
its emission spectrum is very similar to that of a free Rb$_{2}$ molecule.  The interaction between the
elliptical orbital of the $X^{1}\Sigma_{g}$ ground state of Rb$_{2}$ and the attached He atoms is repulsive and
shifts the ground state towards higher energies. It is responsible for the redshift of the measured emission
spectrum whose pressure-independence remains an open question. After fluorescing the exciplex dissociates and the
He atoms return to the bubble interface. The 1042 nm fluorescence thus does not originate from the metastable
(1)$^{3}\Pi_{u}$ state directly, but rather from an exciplex formed by that state.

\subsection{Photodissociation spectrum}

The excitation spectrum for atomic emission at 760 nm is presented in Fig.~\ref{fig:excitationspec}(c). Only the
doublet labelled "h" represents absorption by individual Rb atoms ($D_{1}$ and $D_{2}$ absorption lines). The
peaks: "d", "e", "i", and "j" represent photodissociation lines of the Rb$_{2}$ molecule. Three of them ("d",
"e", and "i") lie very close to the molecular absorption bands discussed in the previous section, while the peak
"j" has no analog in the absorption spectrum of Fig.~\ref{fig:excitationspec}(a).

The calculated Franck-Condon factors for the relevant molecular bands are shown in
Fig.~\ref{fig:excitationspec}(d). The lineshapes of the bound-free transition (1)$^{3}\Sigma_{u} \rightarrow$
(2)$^{3}\Sigma_{u}$ (peak "i") and of the bound-free part of the (1)$^{3}\Sigma_{u} \rightarrow$ (1)$^{3}\Pi_{g}$
band (peak "b") are calculated based on a standard adiabatic line-broadening theory \cite{AllardKielkopf;Review}
as the Fourier transform of the autocorrelation function

\begin{equation}  \label{eq:autocorrelation}
C(\tau)=exp\{-\int(1-exp[-i \Delta\nu(r)\tau])\Psi_{g}^{(0)}(r)^{2}dr\},
\end{equation}
where $\Delta\nu(r)$ represents the transition energy corresponding to the interatomic separation $r$. Only the
bound-free part of the (1)$^{3}\Sigma_{u} \rightarrow$ (1)$^{3}\Pi_{g}$ band is shown in
Fig.~\ref{fig:excitationspec}(d), whereas the bound-bound components of this band are represented in
Fig.~\ref{fig:excitationspec}(b) by their Franck-Condon factors.

The blue component of the atomic absorption doublet "h" ($D_{2}$ resonance line) overlaps with the molecular
(1)$^{3}\Sigma_{u} \rightarrow$ (1)$^{3}\Pi_{g}$ absorption band ("b" in Fig.~\ref{fig:excitationspec}(a)). As
can be seen from Fig.~\ref{fig:potentials}, the position of the minimum (6.1~\AA) of the shallow potential well
of the triplet ground-state is much larger than that of the (1)$^{3}\Pi_{g}$ state. The Condon point of the
excited state is right at the top of the potential barrier. Excitations occurring at smaller distances thus
populate vibrational states which rapidly relax towards the bottom of the potential well with subsequent
radiationless transitions to lower-lying electronic states, and finally  to the (1)$^{3}\Pi_{u}$ state. On the
other hand, absorption at larger internuclear distances populates the right side of the (1)$^{3}\Pi_{g}$
potential barrier which results in the dissociation of the molecule into a 5$S$ ground-state atom and an atom in
the excited 5$P$ state. This dissociation channel cannot be distinguished from the purely atomic
absorption-emission cycle in our experiments. A similar competition between quenching and dissociation of the
(1)$^{3}\Pi_{g}$ state was observed in experiments on potassium dimers attached to helium nanodroplets
\cite{Scoles;AlkaliDimersTripletManifold,Reho;Dynamics3PigStateK2}, where the atomic absorption lines and the
molecular (1)$^{3}\Sigma_{u} \rightarrow$ (1)$^{3}\Pi_{g}$ band were resolved. In those experiments the quenching
of the laser-excited (1)$^{3}\Pi_{g}$ state of K$_{2}$ resulted in fluorescence at the two allowed
singlet-singlet transitions, whose upper states lie below the (1)$^{3}\Pi_{g}$ state: $A^{1}\Sigma_{u}
\rightarrow X^{1}\Sigma_{g}$ and $B^{1}\Pi_{u} \rightarrow X^{1}\Sigma_{g}$. The fast desorption of the excited
molecule from the surface of the droplet prevented the population of the metastable (1)$^{3}\Pi_{g}$ state to be
observed in that study.

The photodissociation band "i" has almost the same wavelength as the $X^{1}\Sigma_{g} \rightarrow B^{1}\Pi_{u}$
band in Fig.~\ref{fig:excitationspec}(a) (labelled "c"). However, the Condon point of the $B^{1}\Pi_{u}$
potential lies well below the dissociation limit, and hence no photodissociation is expected. We therefore
attribute the peak "i" to a dissociation of the dimer following absorption from the triplet ground-state. The
only band that has a corresponding wavelength is the forbidden (1)$^{3}\Sigma_{u} \rightarrow$
(2)$^{3}\Sigma_{u}$ transition at 670 nm. The excited state is strongly repulsive and the shift due to the
interaction with the matrix has the same sign as for all other bands and a magnitude of 20 nm. The
photodissociation at the wavelength close to the $X^{1}\Sigma_{g} \rightarrow B^{1}\Pi_{u}$ band of Rb$_{2}$ was
observed in an experiment with He-nanodroplets \cite{Ernst;TripletStatesRbDimers}. The authors of
\cite{Ernst;TripletStatesRbDimers} interpret it as either the forbidden (1)$^{3}\Sigma_{u} \rightarrow$
(2)$^{3}\Sigma_{u}$ transition that becomes allowed due to the perturbation of the Rb$_{2}$ molecule by the
matrix, or as the dissociation of a Rb$_{3}$ molecule.

The photodissociation peak at 625 nm overlaps with the peak "d" of Fig.~\ref{fig:excitationspec}(a), and
therefore we assign it to the same forbidden (1)$^{3}\Sigma_{u} \rightarrow$ (3)$^{1}\Sigma_{g}$ transition. The
potential curve of the (3)$^{1}\Sigma_{g}$ state below the Condon point is crossed by the repulsive
(2)$^{3}\Sigma_{u}$ state and by the strongly bound (2)$^{3}\Sigma_{g}$ state. The transition to the former
results in the dissociation of the dimer, while the latter is connected by an allowed transition to the
metastable (1)$^{3}\Pi_{u}$ state which fluoresces at 1042 nm.

Using the same arguments we can assign the asymmetric photodissociation band at 590 nm (peak "e" of
Fig.~\ref{fig:excitationspec}(c)) to a superposition of the (1)$^{3}\Sigma_{u} \rightarrow$ (2)$^{3}\Sigma_{g}$
and (1)$^{3}\Sigma_{u} \rightarrow$ (2)$^{3}\Pi_{g}$ transitions. The photodissociation is due to the crossing of
the (2)$^{3}\Sigma_{g}$  state with the repulsive (2)$^{3}\Sigma_{u}$ state at R=5.5~\AA, below the Condon point.

Finally, the band peaked at 493 nm (peak "j" in Fig.~\ref{fig:excitationspec}(c)) produces no molecular
fluorescence but only photodissociation. It can be assigned to one of the three forbidden transitions from the
singlet ground-state, viz.,  $X^{1}\Sigma_{g} \rightarrow$ (3)$^{1}\Sigma_{g}$ (505 nm), $X^{1}\Sigma_{g}
\rightarrow$ (2)$^{3}\Pi_{u}$ (496 nm), or $X^{1}\Sigma_{g} \rightarrow$ (2)$^{3}\Sigma_{u}$ (488 nm). The upper
state of the last transition is repulsive, while the upper states of the first two transitions cross it below
their Condon points and may thus also populate it.

\subsection{Photodissociation vs molecular emission at the phase transition}

Here we suggest a speculative qualitative interpretation of the changes in the fluorescence spectrum at the
solid-liquid phase transition. The fluorescence at 1042 nm following the excitation at 650 nm disappears during
the transition to the liquid phase because the upper $B^{1}\Pi_{u}$ state becomes less perturbed by the
surrounding helium and the rate of radiationless transition to the metastable (1)$^{3}\Pi_{u}$ state drops below
that of the allowed radiative transition $B^{1}\Pi_{u} \rightarrow X^{1}\Sigma_{g}$. This emission is difficult
to detect experimentally because of a very strong background from scattered laser light at 650 nm. Nonetheless
the red wing of this band appears in the experimental spectra close to the atomic doublet. On the other hand, the
photodissociation at 650 nm is much less sensitive to changes in the helium density because the dissociating
state is populated directly by absorption and involves no radiationless transition.

The absorption at 580 nm populates the states (2)$^{3}\Sigma_{g}$ and (2)$^{3}\Pi_{g}$ which themselves do not
dissociate but which have allowed radiative transitions to the metastable (1)$^{3}\Pi_{u}$ state and thus do
fluoresce at 1042\,nm. In consequence, a reduction of the helium density results in a suppression of the
photodissociation due to matrix induced mixing with dissociative states, while the metastable state is still
efficiently populated and thus emits fluorescence.

\begin{figure}[tbp]
\includegraphics[width=6. cm]{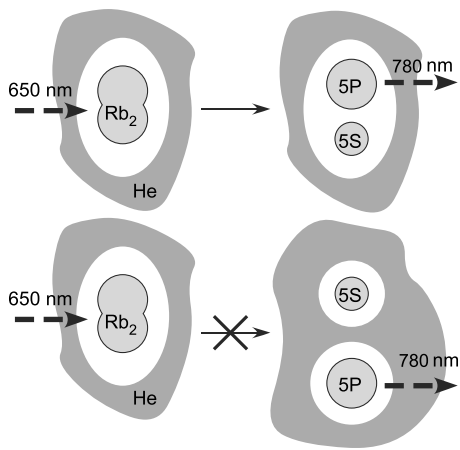}
\caption{Possible final bubble configurations following
dissociation of the Rb$_2$ dimer by absorption of 650\,nm light.
The situation shown on top is energetically favored. Experimental
evidence that the upper graph describes the dissociation process
comes from the fact that the atomic fluorescence line emitted
after photodissociation is less shifted than the well studied
emission from the atomic bubble (bottom graph). }
\label{fig:molbubble}
\end{figure}

The systematic difference of the spectral shifts of the atomic emission lines following excitation via different
channels (atomic absorption at 760 nm or photodissociation at 650 nm) suggests that the presence of the second
(ground-state) Rb atom (resulting from the dissociation) reduces the perturbation imposed by the surrounding He
on the fluorescence emitting Rb atom. The bubble model
\cite{Kinoshita;OpitcalPropAlkaliAtomsPressSuperfluidHe,Kanorsky;PressureshiftHelium} successfully predicts the
shifts of the resonance lines of Rb and Cs bubbles in liquid and solid He. As the electronic density of the
excited $P$ state in the alkali occupies a larger volume than its ground state, it has a larger overlap with the
surrounding He. Due to the Pauli repulsion, this overlap results in a shift of the atomic levels towards higher
energies. Because of the larger shift of the excited state the transition becomes blue shifted.

Based on the bubble model we suggest the following qualitative interpretation of the observed lineshifts at the
phase transition. The Rb$_2$ dimer forms a molecular bubble whose shape reflects the symmetry of the molecule's
electronic configuration. The photodissociation of Rb$_2$ does not result in the formation of two independent
atoms occupying each an individual bubble, but the two Rb atoms, separated by several \AA, will rather reside in
a single large helium bubble (Fig.\,\ref{fig:molbubble}). It is not clear a priori whether this diatomic bubble
configuration is energetically favored over the situation, in which atoms reside in individual bubbles. Because
of the larger average distance of the first He solvation shell this molecular bubble will perturb the atomic
levels to a lesser extent than the smaller monoatomic bubble (studied in
\cite{Kinoshita;OpitcalPropAlkaliAtomsPressSuperfluidHe,Kanorsky;PressureshiftHelium}) and could therefore
explain the smaller spectral shifts of the observed emission lines.

\section{Summary\label{seq:summary}}

We have presented results of laser-induced fluorescence
experiments in solid He doped by laser ablation from a metallic
rubidium target. We have observed for the first time fluorescence
from Rb atoms and from Rb$_2$ dimers in solid He. Atomic emission
was observed after atomic excitation as well as after excitation
of photodissociating states in the Rb$_2$ molecule. In the latter
case the atomic emission lines are less perturbed than after
direct atomic excitation, a fact which we attribute to emission
from diatomic bubbles.

Absorption and emission bands of Rb$_2$ were studied in the visible and near infrared domains. All absorption and
photodissociation bands in this range could be assigned by comparison with calculated Franck-Condon factors. No
emission from any of the allowed transitions of the free Rb$_2$ molecule could be observed in the investigated
spectral range. At the same time we found a large number of absorption bands which all lead to emission of
fluorescence on the forbidden transition from the metastable (1)$^{3}\Pi_{u}$ state to the ground state. The
emitting state was found to have a lifetime of 45 $\mu$s and a formation time of 15 $\mu$s. We interpret this
feature by proposing that the metastable state forms (on a time scale of 15 $\mu$s) a
Rb$_{2}(^{3}\Pi_{u})$He$_{N}$ molecular exciplex, which then decays with a lifetime of 45 $\mu$s.

%\appendix*

\begin{acknowledgments}
This work was supported by grant No.~200020--103864 of the
Schweizerischer Nationalfonds.
\end{acknowledgments}

\bibliography{Rbdimer}

\end{document}